\documentclass[pra,showpacs,preprintnumbers,superscriptaddress,amsmath,amssymb,twocolumn,floatfix]{revtex4}
\usepackage{graphicx}
\usepackage{dcolumn}
\usepackage{bm}
\usepackage{psfrag}
\usepackage{epsfig}
\usepackage{amsmath}
\usepackage{amssymb}
\usepackage{color}
\usepackage{mathbbol}
\usepackage{subfigure}

\newcommand{\nn}{\nonumber}

\begin{document}

\title{Quantum light by atomic arrays in optical resonators }

\author{Hessam Habibian}

\affiliation{Grup d'\'{O}ptica, Departament de F\'{i}sica,
Universitat Aut\`{o}noma de Barcelona, E-08193 Bellaterra,
Barcelona, Spain}
\affiliation{Theoretische Physik, Universit\"{a}t des Saarlandes,
D-66041 Saarbr\"{u}cken, Germany}

\author{Stefano Zippilli}

\affiliation{Grup d'\'{O}ptica, Departament de F\'{i}sica,
Universitat Aut\`{o}noma de Barcelona, E-08193 Bellaterra,
Barcelona, Spain}
\affiliation{Theoretische Physik, Universit\"{a}t des Saarlandes,
D-66041 Saarbr\"{u}cken, Germany}
\affiliation{Department of Physics,
Technische Universit\"{a}t Kaiserslautern, D-67663 Kaiserslautern, Germany}

\author{Giovanna Morigi}

\affiliation{Grup d'\'{O}ptica, Departament de F\'{i}sica,
Universitat Aut\`{o}noma de Barcelona, E-08193 Bellaterra,
Barcelona, Spain}
\affiliation{Theoretische Physik, Universit\"{a}t des Saarlandes,
D-66041 Saarbr\"{u}cken, Germany}

\date{\today}

\begin{abstract}
Light scattering by a periodic atomic array is studied when the atoms couple with the mode of a high-finesse optical resonator and are driven by a laser. When the von-Laue condition is not satified, there is no coherent emission into the cavity mode, and the latter is pumped via inelastic scattering processes. We consider this situation and identify conditions for which different non-linear optical processes can occur. We show that these processes can be controlled by suitably tuning the strength of laser and cavity coupling, the angle between laser and cavity axis, and the array periodicity. We characterize the coherence properties of the light when the system can either operate as degenerate parametric amplifier or as a source of antibunched-light. Our study permits us to identify the individual multi-photon components of the nonlinear optical response of the atomic array and the corresponding parameter regimes, thereby in principle allowing one for controlling the nonlinear optical response of the medium.
\end{abstract}

\pacs{42.50.Nn, 42.50.Pq, 42.65.Yj}

\maketitle

\section{Introduction}

Resonance fluorescence from a single atom exhibits non-classical features~\cite{Resonance-Fluorescence}, which become evident in the correlation functions of the emitted light~\cite{MandelWolf}. Non-classical properties emerge from the quantum nature of the scatterer, such as the discrete spectrum of the electronic bound states of the scattering atom. They can be enhanced or suppressed by several scatterers forming a regular array~\cite{Mandel,Skornia,Vogel85}. In this case, at the solid angles which satisfy the von-Laue condition~\cite{Ashcroft}, the light in the far field is in a squeezed coherent state, while for a large number of atoms it can exhibit vacuum squeezing at scattering angles, for which the elastic component of the scattered light is suppressed~\cite{Vogel85}.

When the atoms of the array are strongly coupled with the mode of a high-finesse resonator, emission into the cavity mode is in general expected to be enhanced. The properties of the light at the cavity output will depend on the phase-matching conditions, determined by the angle between laser and cavity wave vector and by the periodicity of the atomic array. The coherence properties of the light at the cavity output may however be significantly different from the ones predicted in free space. An interesting example is found when the geometry of the setup is such that the atoms coherently scatter light into the cavity mode. In this case the intracavity-field intensity becomes independent of the number of atoms $N$ as $N$ increases, while inelastic scattering is suppressed over the whole solid angle in leading order in $1/N$~\cite{Zippilli_PRL04}. These dynamics have been confirmed by experimental observations~\cite{Black03,Slama}, and clearly differ from the behaviour in free space~\cite{Vogel85}.

When the geometry of the setup is such that the von-Laue condition is not satisfied, photons can only be inelastically scattered into the cavity mode. The smaller system size for which coherent scattering is suppressed is found for two atoms inside the resonator. The properties of the light at the cavity output for this specific case have been studied in Refs.~\cite{Fernandez_07,Evers}. To the best of our knowledge, however, the scaling of the dynamics with the number of atoms $N$ is still largely unexplored in this regime.

In this article we characterize the coherence properties of the light at the cavity output when the light is scattered from a laser into the resonator by an array of atoms and the geometry of the system is such that coherent scattering is suppressed. For the phase-matching conditions, at which in free space the light is in a squeezed-vacuum state~\cite{Vogel85}, we find that inside a resonator and at large $N$ the system behaves as an optical parametric oscillator, which in certain regimes can operate above threshold~\cite{MilburnWalls}. For a small number of atoms $N$, on the contrary, the medium can act as a source of antibunched light. In this case it can either behave as single-photon or, for the saturation parameters here considered, two-photon ``gateway''~\cite{Kubanek08}. The latter behaviour is found for a specific phase-matching condition. We identify the parameter regimes which allow one to control the specific nonlinear optical response of the medium.

This article is organized as follows. In Sec.~\ref{Sec:2} the theoretical model is introduced and the basic approximations are discussed, which lead to the derivation of an effective Hamiltonian for the atomic and cavity excitations. In Sec.~\ref{Sec:3} we analyze the light at the cavity output under the condition that there is no coherent scattering into the cavity. The conclusions are drawn and outlooks are discussed in Sec.~\ref{Sec:5}.

 \section{Theoretical Model}
\label{Sec:2}

\begin{figure}
\centering
\subfigure[]{
\includegraphics[width=7cm]{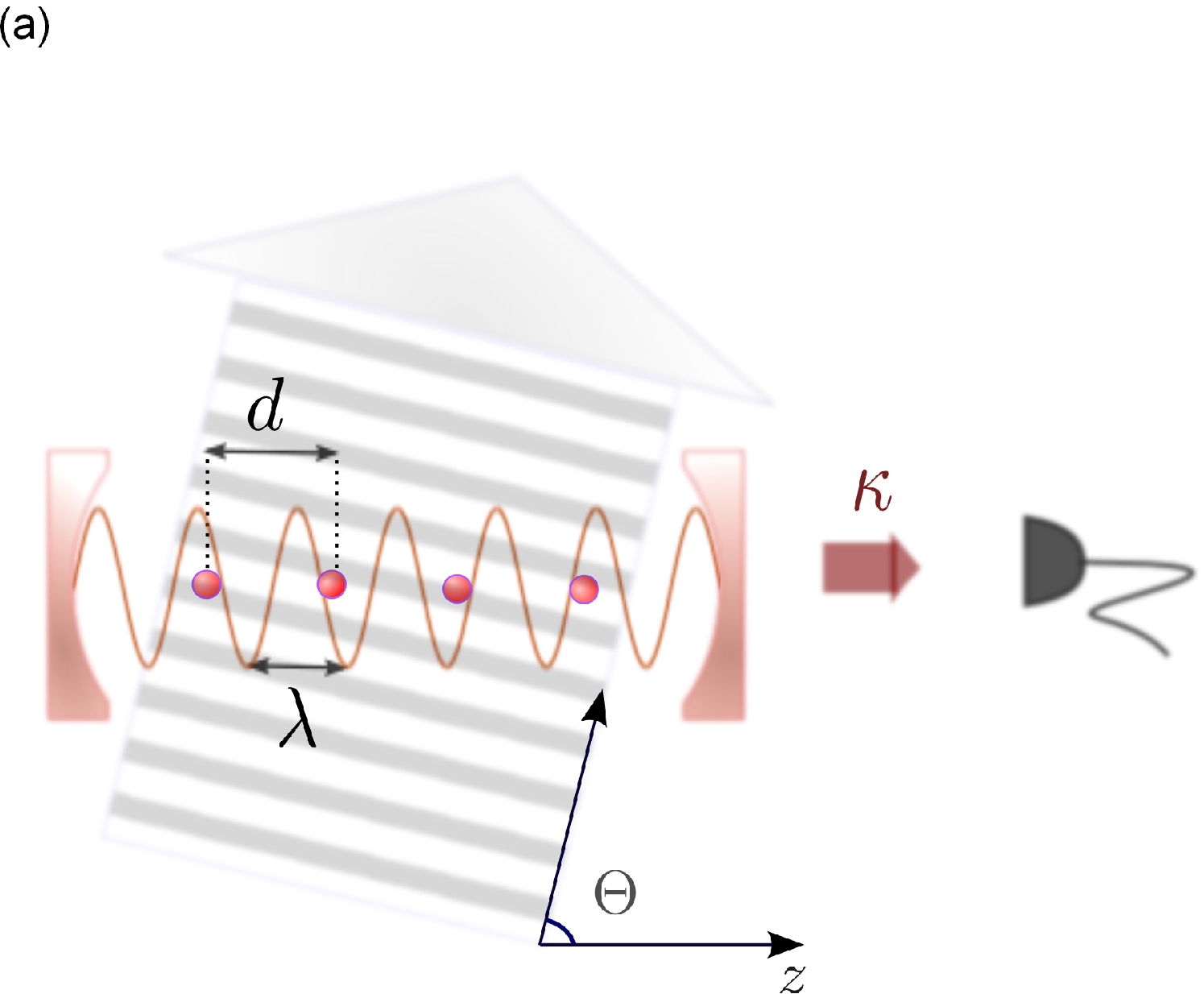}}
\hspace{1.5cm}
\subfigure[]{
\includegraphics[width=4cm]{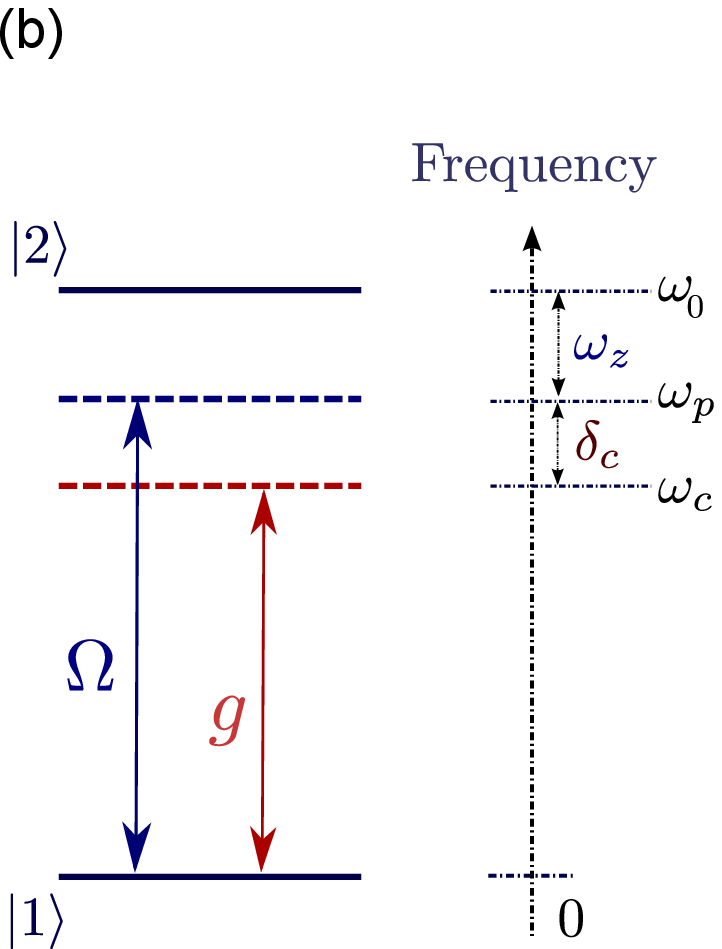}
}
\caption[]{(Color online) (a) An array of atoms, with interparticle distance $d$, is confined along the axis of a standing-wave optical cavity at frequency $\omega_c$ and is transversally driven by a laser, whose wave vector forms the angle $\Theta$ with the cavity axis. The atomic internal transition and the relevant frequency scales are given in (b), with $|1\rangle$ and $|2\rangle$ ground and excited state of an optical transition with frequency $\omega_0$ and natural linewidth $\gamma$. The frequencies $\omega_z=\omega_0-\omega_p$ and $\delta_c=\omega_c-\omega_p$ denote the detunings between the laser frequency $\omega_p$ and the atomic and cavity frequency, respectively. The other parameters are the laser Rabi frequency $\Omega$, the atom-cavity coupling strength $g$, and the decay rate $\kappa$ of the optical cavity. }
\label{Fig:lattice}
\end{figure}

The physical system is composed by $N$ identical atoms which are regularly distributed along the $z$ axis. They are located at the positions $z_j=jd$ where $j=1,\dots,N$ and $d$ is the interparticle distance~\cite{Footnote:0}. An optical dipole transition of the confined atoms interacts with the mode of a standing wave cavity, whose wave vector $k$ is parallel to the atomic array, as illustrated in Fig.~\ref{Fig:lattice}. Moreover, the atoms are transversally driven by a laser and scatter photons into the cavity mode. We denote by $\omega_p$ the frequency of the laser mode, whose polarization is assumed to be linear and parallel to the polarization of the cavity mode, and $a$ and $a^{\dagger}$ the annihilation and creation operators, respectively, of a cavity photon at frequency $\omega_c$ (with $[a,a^{\dagger}]=1$). Cavity and laser mode couple to the atomic dipolar transition at frequency $\omega_0$ with ground and excited states $|1\rangle$ and $|2\rangle$. The Hamiltonian governing the coherent dynamics of cavity mode and the atoms is given by
\begin{align}\label{fullHamil}
\mathcal{H}=&\hbar\omega_c a^{\dag}a+\hbar\omega_0\sum_{j=1}^{N}
S_j^z\nn\\
&+\hbar g\sum_{j=1}^{N}
\cos{(kz_j+\varphi)}(S_j^{\dag}a+a^{\dag}S_j)\nn\\
&+i\hbar\Omega\sum_{j=1}^{N}(S_j^{\dag}e^{-{\rm i}\omega_p t}
e^{{\rm i}(k_pz_j\cos\Theta-\phi_L)}-{\rm H.c.})\,,
\end{align}
where $\Omega$ is the strength of the coupling between laser and atomic transition, and $g$ the cavity vacuum Rabi frequency. The operators $S_j=|1\rangle_j\langle2|$ and $S_j^\dag$ indicate the lowering and raising operators for the atom
at the position $z_j$,  and $S_j^z=\frac{1}{2}(|2\rangle_j\langle2|-|1\rangle_j\langle1|)$ is the $z$ component of the pseudo-spin operator.  In Eq.~(\ref{fullHamil}) we have introduced the angle $\varphi$, which is the phase offset of the standing wave at the atomic positions, the phase of the laser $\phi_L$, and the angle $\Theta$ between the laser and the cavity wave vector.

We remark that in the present study we neglect the atomic motion, and consider that the size of the atomic wavepacket is much smaller than the laser wavelength and interparticle distance. We refer the reader to \cite{Fernandez_07} for a quantitative study on the effects of the mechanical motion on the nonlinear optical processes in this kind of system.

In the rest of this section we will introduce and discuss the approximations, which allow us to solve the dynamics and determine the properties of the cavity field. For simplicity we also set $k=k_p$: The difference between the laser and cavity wave numbers can in fact be neglected for the purpose of this study.

\subsection{Weak excitation limit}

We resort to the Holstein-Primakoff representation for the spin operator~\cite{holstein}
\begin{eqnarray}\label{HPnapp}
&&S_j^{\dag}=b_j^{\dag}(1-b_j^{\dag}b_j)^{1/2}\,,\\
&&S_j=(1-b_j^{\dag}b_j)^{1/2}b_j,\nn\\
&&S_j^z=b_j^{\dag}b_j-\frac{1}{2},\nn
\end{eqnarray}
where $b_j$ ($b_j^\dagger$) is the bosonic operator annihilating (creating) an excitation of the atom at $z_j$, such that $[b_j,b^\dag_{j'}]=\delta_{jj'}$. In the limit in which the atomic dipoles are driven below saturation, we treat saturation effects in the lowest non-vanishing order of a perturbative expansion, whose small parameter is the total excited-state population of the atoms, denoted by ${\cal N}_{\rm tot}$. We denote the detuning of the laser from the atomic transition by
\begin{equation}
\omega_z=\omega_0-\omega_p\,.
\end{equation}
and by $\gamma$ the natural linewidth. In the low saturation limit, $|\omega_z+{\rm i}\gamma/2|\gg\sqrt{N}\Omega$, then ${\cal N}_{\rm tot}\ll N$ and we can expand the operators on the right-hand side of the equations~(\ref{HPnapp}) in second order in the small parameter $\langle b_j^{\dagger}b_j\rangle \ll 1$, obtaining
\begin{eqnarray}\label{HP1}
&&S_j^{\dag}\approx b_j^{\dag}
-\frac{1}{2}b_j^{\dag}b_j^{\dag}b_j\,,\\
&&S_j\approx b_j-\frac{1}{2}b_j^{\dag}b_j b_j\,.
\end{eqnarray}
For $N\gg 1$ the dynamics is expected to be irrelevantly affected by the assumptions on the boundaries. Therefore, we take periodic boundary conditions on the lattice, such that $z_{N+1}=z_1$. The atomic excitations are studied in the Fourier transformed variable $q$, quasimomentum of the lattice, which is defined in the Brillouin zone (BZ) $q\in(-G_0/2, G_0/2]$ with $G_0=2\pi/d$ the primitive reciprocal lattice vector. Correspondingly, we introduce the operators $b_q$ and $b_q^{\dagger}$, defined as
\begin{eqnarray}
\label{Fourier}
&&b_q=\frac{1}{\sqrt{N}}\sum_{j=1}^N b_j e^{-{\rm i}qjd}\,,\\
&&b_{q}^{\dagger}=\frac{1}{\sqrt{N}}\sum_{j=1}^N b_j^{\dagger} e^{{\rm i}qjd}\,,
\end{eqnarray}
which annihilate and create, respectively, an excitation of the spin wave at quasimomentum $q$ and fulfilling the commutation relation $[b_q,b_{q'}^{\dagger}]=\delta_{q,q'}$. After rewriting the Hamiltonian in Eq.~(\ref{fullHamil}) in terms of spin-wave operators, we find
\begin{equation}\label{Hgeneral}
\mathcal{H}\approx-\frac{N\hbar\omega_z}{2}+H_{\rm pump}+{\cal H}^{(2)}+{\cal H}^{(4)}\,,
\end{equation}
where the first term on the Right-Hand Side (RHS) is a constant and will be discarded from now on, while
\begin{equation}
\label{H:pump}
H_{\rm pump}={\rm i}\hbar\Omega\sqrt{N}\left(b_{Q'}^{\dag}e^{-{\rm i}(\omega_pt+\phi_L)}
-b_{Q'}e^{{\rm i}(\omega_pt+\phi_L)}\right)
\end{equation}
is the linear term describing the coupling with the laser. Term
\begin{align}
{\cal H}^{(2)}=&\hbar\omega_c a^{\dag}a+\hbar\omega_0 \sum_{q\in
BZ}b_q^{\dag}b_q\nn\\
&+\frac{\hbar g\sqrt{N}}{2}\left[(b_{Q}^{\dagger}\;e^{i\varphi}+b_{-Q}^{\dagger}\;e^{-i\varphi})a+{\rm H.c.}\right]\label{Hharm}
\end{align}
determines the system dynamics when the linear pump is set to zero and the dipoles are approximated by harmonic oscillators (analog of the classical model of the elastically bound electron), while
\begin{align}
{\cal H}^{(4)}&=-\frac{\hbar g}{4\sqrt{N}}\sum_{q_1,q_2\in
BZ}(b^{\dag}_{q_1}
b^{\dag}_{q_2}b_{q_1+q_2-Q}a\;e^{i\varphi}\nn\\
&+b^{\dag}_{q_1}b^{\dag}_{q_2}b_{q_1+q_2+Q}a\;e^{-i\varphi}+{\rm H.c.})\nn\\
&-i\frac{\hbar\Omega}{2\sqrt{N}}\sum_{q_1,q_2\in BZ}(b_{q_1}^{\dag}b_{q_2}^{\dag}b_{q_1+q_2-Q'}e^{-{\rm i}(\omega_pt+\phi_L)}-{\rm H.c.})\label{Hnharm}
\end{align}
accounts for the lowest-order corrections due to saturation. In Eqs.~(\ref{Hharm}) and~(\ref{Hnharm}) we have denoted by  $\pm Q$ and $Q'$ the quasimomenta of the spin waves which couple to the cavity and laser mode, respectively, and which fulfill the phase matching conditions
\begin{eqnarray}
&&Q=k+G\,,\label{phase:1}\\
&&Q'=k\cos\Theta +G'\,,\label{phase:2}
\end{eqnarray}
with reciprocal vectors $G,G'$ such that $Q,Q'\in$  BZ. The atoms scatter coherently into the cavity mode when the von-Laue condition is satisfied, namely one of the two relations is fulfilled:
\begin{eqnarray}
\label{vonLaue}
&&2k\sin^2(\Theta/2)=nG_0\,,\\
&&2k\cos^2(\Theta/2)=n'G_0\,,
\end{eqnarray}
with $n,n'$ integer numbers. In free space, the von-Laue condition corresponds to Eq. (\ref{vonLaue}): for these angles one finds squeezed-coherent states in the far field~\cite{Vogel85}. When the scattered mode for which the von-Laue condition is fulfilled corresponds to a cavity mode, superradiant scattering enhances this behaviour, until the number of atoms $N$ is sufficiently large such that the cooperativity exceeds unity. In this limit one observes saturation of the intracavity-field intensity, which reaches an asymptotic value whose amplitude is independent of $N$ as $N$ is further increased. In the limit $N\gg 1$ the light at the cavity output is in a coherent state, while inelastic scattering is suppressed at leading order in $1/N$~\cite{Zippilli_PRL04}.

When the von-Laue condition is not satisfied, classical mechanics predicts that there is no scattering into the cavity mode. These modes of the electromagnetic fields are solely populated by inelastic scattering processes. Moreover, in free space, when
\begin{equation}
\label{Condition:Squeeze}
 2k\sin^2(\Theta/2)=(2n+1)G_0/2\,,
\end{equation}
then the inelastically scattered light is in a vacuum-squeezed state~\cite{Vogel85}. Inside a standing-wave resonator, on the other hand, the mode is in a vacuum-squeezed state provided that either Eq.~(\ref{Condition:Squeeze}) or an additional relation,
\begin{equation}
2k\cos^2(\Theta/2)=(2n+1)G_0/2\,,
\end{equation}
is satisfied.

In the following we will study the field at the cavity output as determined by the dynamics of Hamiltonian~(\ref{Hgeneral}) when $Q'\neq \pm Q$, namely, when the scattering processes which pump the cavity are solely inelastic. We remark that throughout this treatment we do not make specific assumptions about the ratio between the array periodicity $d$ and the light wavelength $\lambda$ (and therefore also consider the situation in which $\lambda\neq 2d$. This situation has been experimentally realized for instance in Refs.~\cite{Kimble,Kinner,Slama,Drewsen}).

\subsection{Linear response: Polaritonic modes}

We first solve the dynamics governed by Hamiltonian $\mathcal H^{(2)}$ in Eq.~(\ref{Hharm}). In the diagonal form the quadratic part can be rewritten as
\begin{equation}
\mathcal H^{(2)}=\sum_{j=1}^{2} \hbar\omega_j\gamma_j^{\dag}\gamma_j +\sum_{q\neq Q_s,q\in BZ} \hbar\omega_0 b_q^{\dag}b_q\,,
\end{equation}
where $Q_s$ labels the spin wave which couples with the cavity mode, such that
\begin{eqnarray}
\label{Sym:Polariton}
&&b_{Q_s}=b_Q \mbox{~~if~~}Q=0,G_0/2\,, \\ &&b_{Q_s}=\frac{b_{Q}\;e^{-i\varphi}+b_{-Q}\;e^{i\varphi}}{\sqrt{2}}\mbox{~~otherwise}\,.\label{Sym:Polariton:1}
\end{eqnarray}
The resulting polaritonic eigenmodes are \begin{align}
&\gamma_1=-a\cos X+b_{Q_s}\sin X\,, \label{polariton:1}\\
&\gamma_2=a\sin X+b_{Q_s}\cos X\,,\label{polariton:2}
\end{align}
with respective eigenfrequencies \begin{eqnarray}\label{EigenFreqs}
&&\omega_{1,2} = \frac{1}{2}\left(\omega_c+\omega_0 \mp \delta\omega\right),\\
&&\delta\omega=\sqrt{\left(\omega_0-\omega_c\right)^2+4\tilde{g}^2N}\;\,,
\end{eqnarray}
and
\begin{equation}
\label{angle:X}
\tan X=\tilde{g}\sqrt{N}/(\omega_0-\omega_1)\,,
\end{equation}
which defines the mixing angle $X$. The parameter $\tilde{g}$ is proportional to the coupling strength. In particular, $\tilde{g}=g\cos\varphi$ when $Q=0,G_0/2$ and the cavity mode couples with the spin wave $b_{Q_s}=b_Q$, otherwise $\tilde{g}=g/\sqrt{2}$ and the spin wave is given in Eq.~(\ref{Sym:Polariton:1}).

Hamiltonian~(\ref{H:pump}) describes the coupling of the pump with the spin wave  $Q'$. When $Q'\neq \pm Q$, photons are pumped into the cavity via inelastic processes, which in our model are accounted for by the Hamiltonian term in Eq.~(\ref{Hnharm}). On the other had, when the dynamics is considered up to the quadratic term (hence, inelastic processes are neglected), only the mode $Q'$ is pumped and the Heisenberg equation of motion for $b_{Q'}$ reads
\begin{equation}
\label{drive:0}
\dot{b}_{Q'}=-{\rm i}\omega_zb_{Q'}-\frac{\gamma}{2}b_{Q'}+\Omega\sqrt{N}{\rm e}^{-{\rm i}\phi_L}+\sqrt{\gamma}b_{q, \rm in}(t)\,,
\end{equation}
that has been written in the reference frame rotating at the laser frequency $\omega_p$. Here, $\gamma$ is the spontaneous decay rate and $b_{q,\rm in}(t)$ is the corresponding Langevin force operator, such that $\langle b_{q,\rm in}(t)\rangle=0$ and $\langle b_{q,\rm in}(t)b_{q,\rm in}^{\dagger}(t')\rangle=\delta(t-t')$~\cite{MilburnWalls}. The general solution reduces, in the limit in which $|\omega_z|\gg\gamma$, to the form
\begin{equation}
\label{drive}
b_{Q'}\simeq -{\rm i}\frac{\Omega\sqrt{N}}{\omega_z}{\rm e}^{-{\rm i}\phi_L}
\end{equation}
which is consistent with the expansion to lowest order in Eq.~(\ref{HP1}) provided that $\Omega^2 N\ll \omega_z^2$. In the reference frame rotating at the laser frequency the explicit frequency dependence of the Hamiltonian terms is dropped, and $\omega_1\to \omega_1-\omega_p$, $\omega_2\to \omega_2-\omega_p$, $\omega_0\to \omega_z$, and $\omega_c\to \omega_c-\omega_p\equiv \delta_c$.

\subsection{Effective Hamiltonian}\label{sub1}

Under the assumptions discussed so far, we derive from Hamiltonian~(\ref{Hgeneral}) an effective Hamiltonian for the polariton $\gamma_1$. The effective Hamiltonian is obtained by adiabatically eliminating the coupling with the other polaritons, according to the procedure sketched in Appendix A, and reads
\begin{align}\label{Heff_eq:0}
{\cal H}_{\rm eff}=&\hbar\delta\omega_1\gamma^{\dag}_{1}\gamma_{1}\nn\\
&+\frac{\hbar}{2}\big(\alpha\gamma^{\dag 2}_1e^{2i\phi_L}+\alpha^\ast\gamma_1^2e^{-2i\phi_L}\big)
+\hbar\chi\gamma^{\dag}_1\gamma^{\dag}_1\gamma_1\gamma_1\nn\\
&+i\hbar(\nu\gamma_1^{\dag2}\gamma_1e^{i\phi_L}-\nu^*\gamma_1^\dag\gamma_1^2e^{-i\phi_L})\,,
\end{align}
where~\cite{Footnote:1}
\begin{eqnarray}
&&\delta\omega_1=\omega_1-\omega_p+\frac{2\Omega^2}{\omega_z}\left(\tilde{S}^2+\frac{\tilde{g} \sqrt{N}}{\omega_z}\tilde{S}\tilde{C}\right)\,,\label{H:shift}\\
&&\alpha=-\frac{\Omega^2}{\omega_z}\big(\tilde{S}^2+\frac{\tilde{g} \sqrt{N}}{\omega_z}\tilde{S}\tilde{C}\big)
\left(\delta_{Q',G/2}+\mathcal C_{\alpha}^{k\neq G/2}\right)\,,\label{H:alpha}\\
&&\chi=\frac{\tilde{g}}{\sqrt N}\tilde S^3\tilde C\left[1+ \mathcal C_{\chi}^{k\neq G/2}\right],\label{H:chi}\\
&&\nu= -\frac{\Omega}{4\sqrt{N}}\left(\tilde S^3+\frac{3\tilde g\sqrt{N}}{\omega_z}\tilde{S}^2\tilde{C}\right)\mathcal C_{\nu}^{k\neq G/2}\,,\label{H:nu}
\end{eqnarray}
with $\tilde{S}=\sin X$ and $\tilde{C}=\cos X$. The terms $\mathcal C_j^{\;k\neq G/2}$ do not vanish when $k\neq G/2$, and their explicit form is
\begin{eqnarray}
&\mathcal C_\chi^{k\neq G/2}=&\left(\frac{1}{2}+\frac{1}{2}\delta_{Q,\pm G_0/4}\cos{(4\varphi)}\right) (1-\delta_{k,G/2}),\nn\\
&\mathcal C_\alpha^{k\neq G/2}=&\frac{1}{2}\left(\delta_{Q',Q+G/2}e^{-2i\varphi}+\delta_{Q',-Q+G/2}e^{2i\varphi}\right)\nn\\
&& \times(1-\delta_{k,G/2}),\nn\\
&\mathcal C_\nu^{k\neq G/2} =&\frac{1}{\sqrt2}\Big(\delta_{Q',3Q}e^{-3i\varphi}+\delta_{Q',-3Q}e^{3i\varphi}\Big)(1-\delta_{k,G/2})
\nn\,.
\end{eqnarray}
The coefficients have been evaluated under the requirement $Q'\neq \pm Q$.

We now comment on the condition $\Omega^2 N\ll \omega_z^2$, on which the validity of Eq.~(\ref{drive}) is based. When this is not fulfilled, such that $|\langle b_{Q'}\rangle| \sim 1$, Eq.~(\ref{drive:0}) must contain further non-anharmonic terms from the expansion of Eq.~(\ref{HPnapp}) and which account for the saturation effects in $b_{Q'}$. Since this spin mode is weakly coupled to the other modes, which are initially empty, we expect that the polaritons $\gamma_1$ and $\gamma_2$ will remain weakly populated and the structure of their effective Hamiltonian will qualitatively not change.

We also note that in the resonant case, when $\omega_z=0$, the form of Hamiltonian in Eq. (\ref{Heff_eq:0}) remains unchanged, while in the coefficients $\delta\omega_1$, $\alpha$, $\nu$, $\chi$ the following substitution $\omega_z\to (\omega_z^2+\gamma^2/4)/\omega_z$ must be performed . A first consequence is that $\alpha=0$, which implies that there are no processes in this order for which polaritons are created (annihilated) in pairs. A further consequence is that spontaneous decay plays a prominent role in the dynamics. We refer the reader to Sec. \ref{Sec:2:E} for a discussion of the related dissipative effects.

\subsection{Discussion}

\begin{figure}
\centering
\subfigure[]{
\includegraphics[width=7cm]{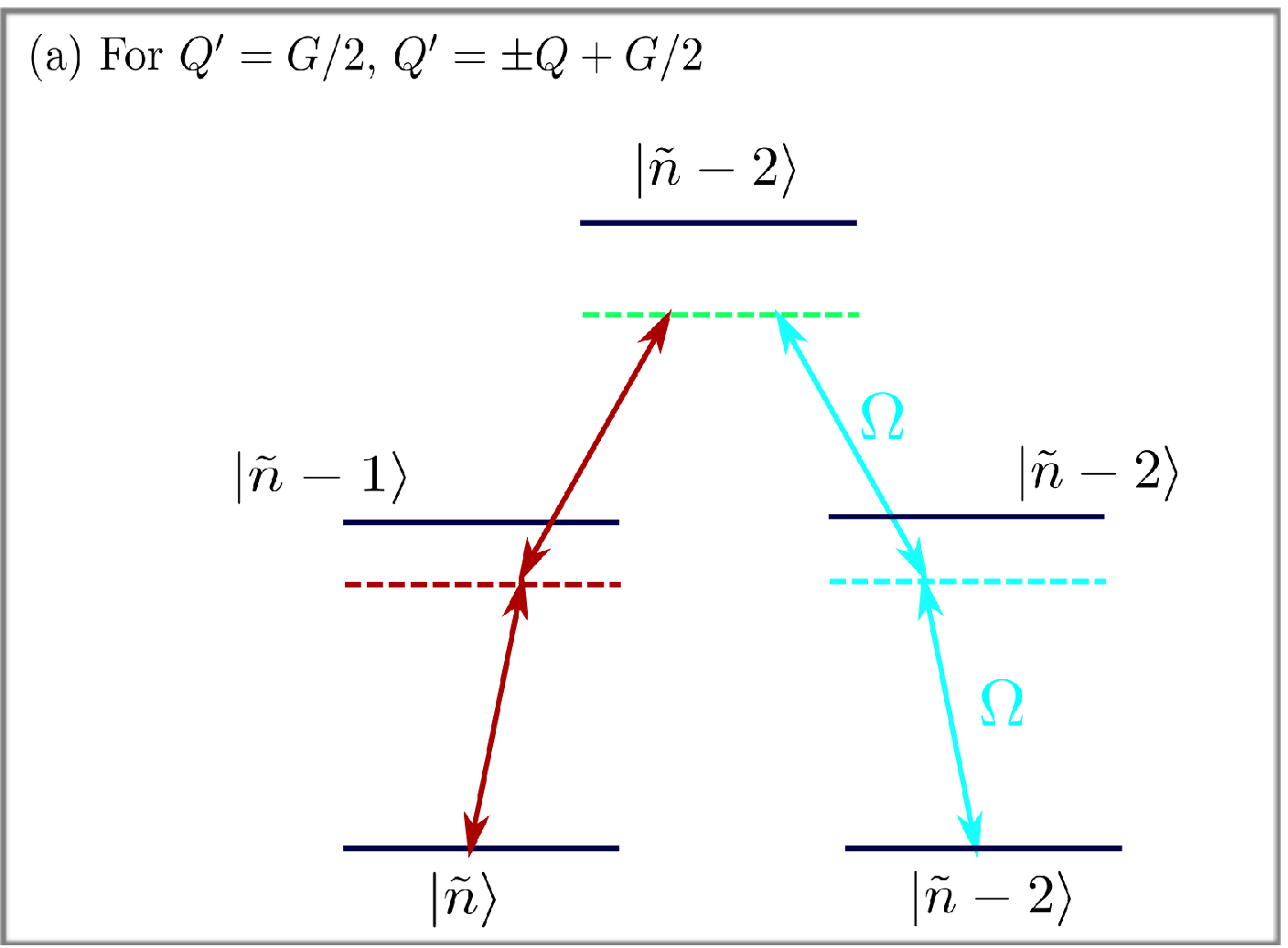}\label{Fig:NLproc_alpha}
}
\subfigure[]{
\includegraphics[width=7cm]{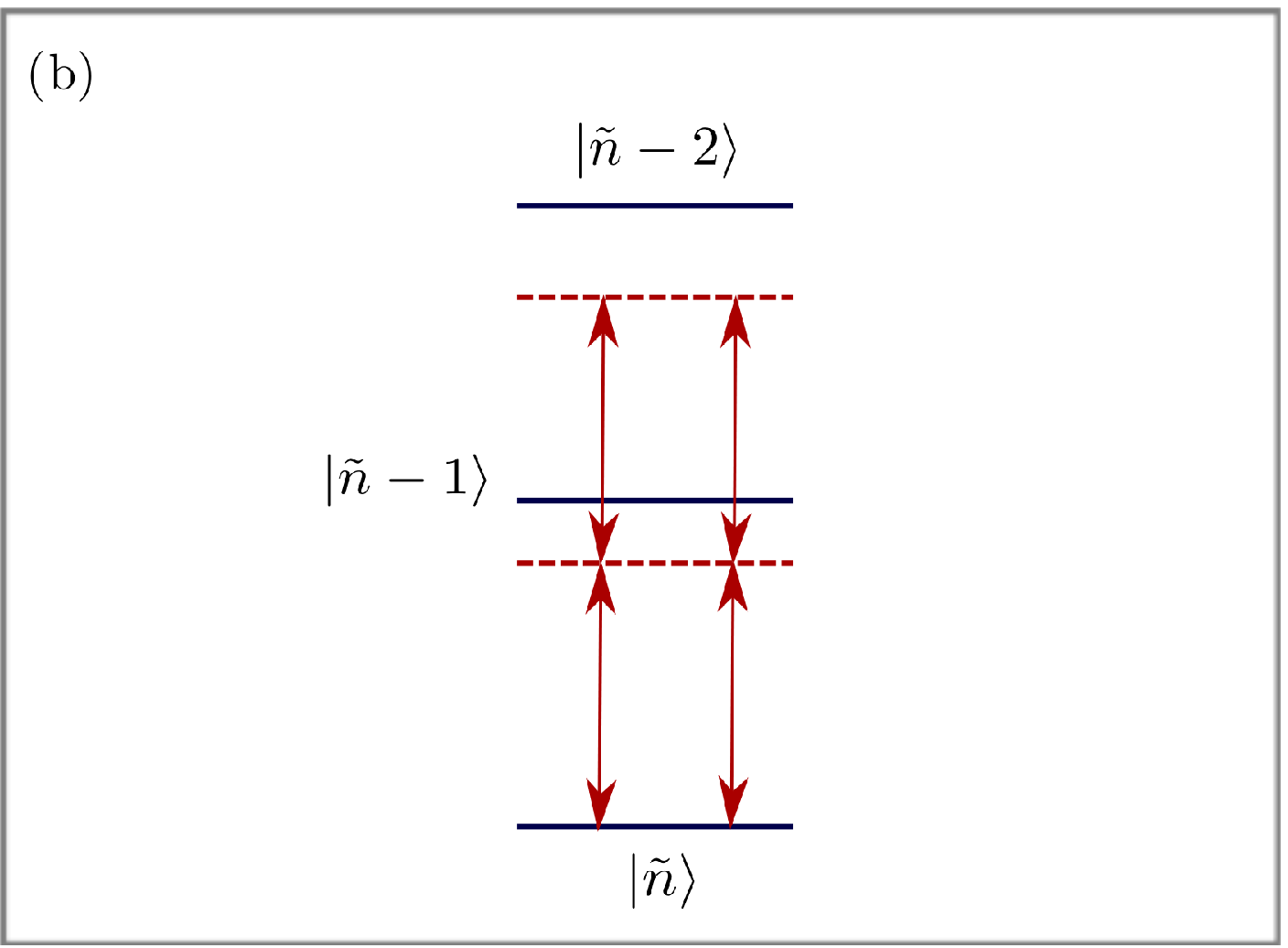}\label{Fig:NLproc_chi}
}
\subfigure[]{
\includegraphics[width=7cm]{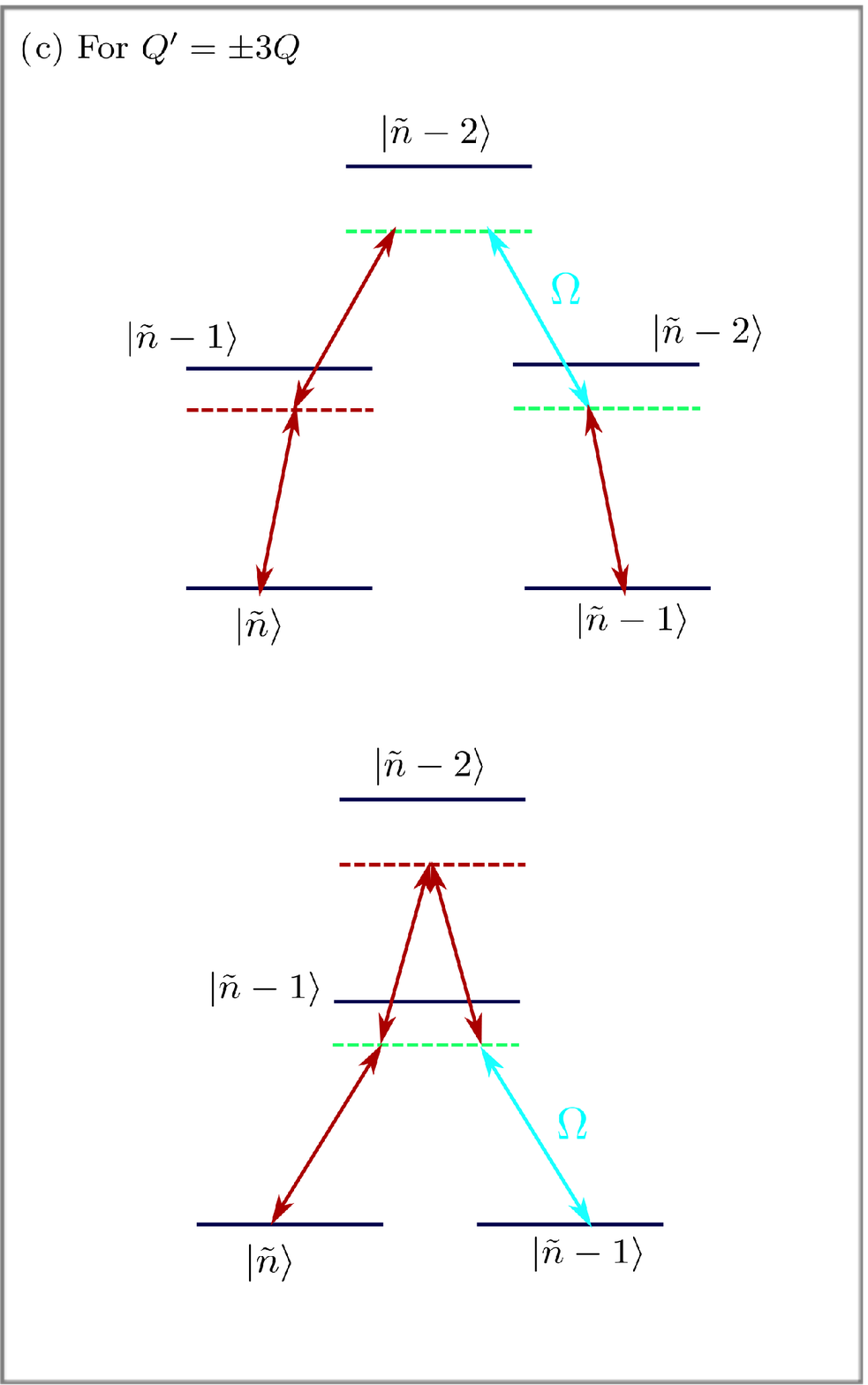}\label{Fig:NLproc_nu}
}
\caption[]{(Color online) Schematic diagram of transitions which fulfill the phase-matching condition till fourth order. State $|\tilde{n}\rangle$ denote the number state of the polariton mode $\gamma_1$. The blue (light gray)
arrows denote the laser-induced couplings and the red (dark gray) arrows denote the creation (annihilation) of polaritons due to the coupling with the cavity field. See text for a detailed discussion.}
\label{Fig:NLproc}
\end{figure}

Let us now discuss the individual terms on the RHS of Eq.~(\ref{Heff_eq:0}). For this purpose it is useful to consider multilevel schemes, which allow one to illustrate the relevant nonlinear processes. The multilevel schemes are depicted in Fig.~\ref{Fig:NLproc}: state $|\tilde{n}\rangle$ is the polariton number state with $\tilde{n}$ excitations. The blue arrows indicate transitions which are coupled by the laser, for which the polariton state is not changed; The red arrows denote transitions which are coupled by the cavity field, for which the polariton state is modified by one excitation.

Using this level scheme, one can explain the dynamical Stark shift $\delta\omega_1$ of the polariton frequency in Eq.~(\ref{H:shift}) as due to higher-order scattering processes, in which laser- and cavity-induced transition creates and then annihilates, in inverse sequential order, a polariton.

The second term on the RHS has coupling strength given in Eq.~(\ref{H:alpha}), it generates squeezing of the polariton and does not vanish provided that $Q'=G/2$ or $Q'=\pm Q+G/2$. The latter condition is equivalent to the free-space condition~(\ref{Condition:Squeeze}), while the first arises from the fact that the cavity mode couples with the symmetric superposition $b_{Q_s}$ in Eq.~(\ref{Sym:Polariton}). The corresponding phase-matched scattering event is a
four-photon process, in which two laser photons are absorbed (emitted) and two polaritonic quanta are  created (annihilated). For $Q'=G/2$ and $Q'\neq \pm Q$ the polaritons are  created in pairs with quasimomentum $Q$ and $-Q$ (the relation $Q=G/2$ corresponds to $b_{-Q}=b_Q$). This specific term is also present when the geometry of the setup is such that von-Laue condition is fulfilled, and at this order is responsible  for the squeezing present in the light at the cavity output.

The Kerr-nonlinearity (third term on the RHS) gives rise to an effective interaction between the polaritons and emerges from processes in which polaritons are absorbed and emitted in pairs. It is depicted in Fig.~\ref{Fig:NLproc_chi} for a generic case. This term is directly proportional to the cavity coupling strength and inversely proportional to $\sqrt{N}$. In this order, it is the term that gives rise to anti-bunching.

The last term on the RHS, finally, is a nonlinear pump of the polariton mode, whose strength depends on the number of polaritonic excitations. It is found when the phase-matching condition $Q'=\pm3Q+G$ is satisfied, which is equivalent to the relation $\cos\theta=(3+n\lambda/d)$ when laser and optical resonator have the same wavelength, as in the case here considered. This relation can be fulfilled for $n\neq 0$ and specific ratios $\lambda/d$. This term vanishes over the vacuum state, and it pumps a polariton at a time with strength proportional to the number of polariton excitations.

In general, photons into the cavity mode are pumped provided that either (i) $Q'=\pm Q$ or (ii) (for $Q'\neq \pm Q$) one of the two conditions are satisfied: $Q'=G/2$ or $Q'=\pm Q+G/2$. We note that the strength of the Rabi frequency and of the cavity Rabi coupling may allow one to tune the relative weight of the various terms in Hamiltonian~(\ref{Heff_eq:0}). Their ratio scales differently with the number of atoms in different regimes, which we will discuss below. Moreover, the interparticle distance of the atomic array constitutes an additional control parameter over the nonlinear optical response of the medium. Further phase-matching conditions are found when considering higher-order terms in the expansion of the spin operators in harmonic-oscillator operators from Eq.~(\ref{HPnapp}). Their role in the dynamics will be relevant, as long as they compete with the dissipative rates, here constituted by the cavity loss rate and spontaneous emission.

\subsection{Cavity input-output formalism}
\label{Sec:2:E}

We consider the full system dynamics, including the atomic spontaneous emission and the cavity quantum noise due to the coupling to the external modes of the electromagnetic field via the finite transmittivity of the cavity mirrors. Denoting by $\kappa$ and $\gamma$ the linewidth of the cavity mode and of the atomic transition, respectively, the Heisenberg-Langevin equations for the operator $a$ and $b_q$ read
\begin{eqnarray}\label{HL:a}
&&\dot{a}=\frac{1}{{\rm i}\hbar}[a,{\mathcal H}]-\kappa a+\sqrt{2\kappa}a_{\rm in}(t)\,,\\
&&\dot{b}_q=\frac{1}{{\rm i}\hbar}[b_q,{\mathcal H}]-\frac{\gamma}{2} b_q+\sqrt{\gamma}b_{q,\rm in}(t)\,,
\end{eqnarray}
where $a_{\rm in}$, $b_{q,\rm in}$ are the Langevin operators, which are decorrelated one from the other and fulfill the relations $\langle a_{\rm in}(t) \rangle=0$ and $\langle a_{\rm in}(t)a_{\rm in}^{\dagger}(t')\rangle=\delta(t-t')$. Here, the average $\langle \cdot\rangle$ is taken over the density matrix at time $t=0$ of the system composed by the atomic spins and by the electromagnetic field. The output field $a_{\rm out}$ at the cavity mirror is given by the relation
\begin{equation}\label{a_out}
a_{\rm out}+a_{\rm in}=\sqrt{2\kappa} a(t)\,.
\end{equation}

Let us now consider the scattering processes occurring in the system. They can be classified into three types: (i) a laser photon can be scattered into the modes of the external electromagnetic field (emf)  by the atoms, without the resonator being pumped in an intermediate time; (ii) a laser photon can be scattered into the cavity mode by the atom and then dissipated by cavity decay; (iii) a laser photon can be scattered into the cavity mode by the atom, then been reabsorbed and emitted into the modes of the external emf. Processes of kind (i) include elastic scattering. They can be the fastest processes, but do not affect the properties of the light at the cavity output. Processes of kind (ii) are the ones which outcouple the intracavity field, but need to be sufficiently slow in order to allow for the build-up of the intracavity field. Processes of kind (iii) are detrimental for the nonlinear optical dynamics we intend to observe, as they introduce additional dissipation (see for instance \cite{Morigi2006,Vitali2006} for an extensive discussion and \cite{Fernandez_07} for a system like the one here considered but composed by two atoms).

Processes (iii), i.e., reabsorption of cavity photons followed by spontaneous emission, can be neglected assuming that the laser and cavity mode are far-off resonance from the atomic transition. In this limit, the cavity is pumped by coherent Raman scattering processes and an effective Heisenberg-Langevin equation for the polariton $\gamma_1$ can be derived assuming that its effective linewidth $\kappa_1=\kappa \cos^2 X+(\gamma/2) \sin^2 X$ fulfilling the inequality $\kappa_1\ll \delta\omega$ (that corresponds to the condition for which the vacuum Rabi splitting is visible in the spectrum of transmission~\cite{Kimble:0,Carmichael_VacuumRabi,Kinner,Kasevich,Drewsen}). We find
\begin{eqnarray}
\label{HL}
\dot{\gamma_1}=\frac{1}{{\rm i}\hbar}[\gamma_1,{\mathcal H}_{\rm eff}]-\kappa_1 \gamma_1+\sqrt{2\kappa}\tilde{C} a_{\rm in}(t)+\sqrt{\gamma}\tilde{S} b_{q,\rm in}(t)\,,
\end{eqnarray}
which determines the dissipative dynamics of the polariton. The field at the cavity output is determined using the solution of the Heisenberg Langevin equation~(\ref{HL}) with Eqs.~(\ref{polariton:1})-(\ref{polariton:2}) in Eq.~(\ref{a_out}). In some calculations, when appropriate we solved the corresponding master equation for the density matrix of the polaritonic modes $\gamma_1$ and $\gamma_2$.

Some remarks are in order at this point. Nonlinear-optical effects in an atomic ensemble, which is resonantly pumped by laser fields, have been studied for instance~\cite{Hafezi}, where the nonlinearity is at the single atom level and is generated by appropriately driving a four-level atomic transitions~\cite{Imamoglu}.

It is important to note, moreover, that Eq. (\ref{HL}) is valid as long as the loss mechanisms occur on a rate which is of the same order, if not smaller, than the inelastic processes. This leads to the requirement that the atom-cavity system be in the strong-coupling regime.

\section{Results}
\label{Sec:3}

We now study the properties of the light at the cavity output as a function of various parameters, assuming that $Q'=G/2$ and that the relations $Q'\neq \pm Q$ and $Q'\neq \pm 3Q+G$ hold. Under these conditions the effective Hamiltonian in Eq.~(\ref{Heff_eq:0}) contains solely the squeezing and the Kerr-nonlinearity terms, while $\nu=0$. Moreover, we assume the condition $\kappa_1\simeq \kappa>\gamma$.

The possible regimes which may be encountered can be classified according to whether the ratio $$\varepsilon=|\alpha/\chi|$$ is larger or smaller than unity. In the first case the medium response is essentially the one of a parametric amplifier. In the second case the Kerr non-linearity dominates, and polaritons can only be pumped in pairs provided that the emission of two polaritons is
a resonant process.

Let us now focus on the regime in which the system acts as a parametric amplifier, namely, $\varepsilon\gg 1$. In this case one finds that the number of photons at the cavity output at time $t$ is
$$\langle a_{out}^\dagger a_{out}\rangle_t\simeq 2\kappa \tilde{C}^2\langle \gamma_1^\dagger\gamma_1\rangle_t\,,$$
with
\begin{eqnarray}
\langle \gamma_1^\dagger\gamma_1\rangle_t&=&\frac{1}{2}\frac{\alpha^2}{\kappa_1^2-\alpha^2}+{\rm e}^{-2\kappa_1t}\sinh^2(\alpha t)\\
& &+\frac{{\rm e}^{-2\kappa_1t}}{2}\left(1-\kappa_1\frac{\kappa_1\cosh(2\alpha t)+\alpha\sinh(2\alpha t)}{\kappa_1^2-\alpha^2}\right)\,.\nonumber
\end{eqnarray}
Depending on whether $\alpha>\kappa_1$ or $\alpha<\kappa_1$, one finds that the dynamics of the intracavity polariton corresponds to a parametric oscillator above or below threshold, respectively. In the following we focus on the case below threshold and evaluate the spectrum of squeezing. We first observe that the quadrature $x^{(\theta)}=\gamma_1e^{-i\theta}+\gamma_1^\dag e^{i\theta}$ has minimum variance for $\theta=\pi/4$ and reads~\cite{Fernandez_07}
\begin{equation}
\langle \Delta x^{(\frac{\pi}{4})^2} \rangle_{st}=
\frac{{\kappa}_1}{{\kappa}_1+|{\alpha}|}\,,
\end{equation}
where the subscript $st$ refers to the expectation value taken over the steady-state density matrix. The squeezing spectrum of the maximally squeezed quadrature is
\begin{align}\label{Squeez_spec:0}
S_{out}(\omega)=&
1+\int_{-\infty}^{+\infty}\langle:x_{out}^{(\frac{\pi}{4})}(t+\tau),x_{out}^{(\frac{\pi}{4})}(t)
:\rangle_{st}e^{-{\rm i}\omega
\tau}d\tau \\
=&1-\frac{4{\kappa}\tilde{C}^2
|{\alpha}|}{({\kappa}_1+|{\alpha}|)^2+\omega^2}\,,\label{Squeez_spec}
\end{align}
where $\langle:\;:\rangle_{st}$ indicates the expectation value for the normally-ordered operators over the steady state, with
$$x_{out}^{(\theta)}=a_{\rm out}e^{-i\theta}+a_{\rm out}^\dag e^{i\theta}\,,$$
and $\langle A,B \rangle_{st}=\langle AB\rangle_{st}-\langle A\rangle_{st}\langle B\rangle_{st}$.

We now discuss the parameter regime in which these dynamics can be encountered. The relation $\varepsilon\gg 1$ is found provided that $\Omega\gg g$. When $|\omega_z|\gg \Omega\sqrt{N}$, in this limit $|\alpha|\simeq \Omega^2 g^2 N/\omega_z^3$, and squeezing can be observed only for very small values of $\kappa$. Far less demanding parameter regimes can be accessed when relaxing the condition on the laser Rabi frequency, and assuming that $\Omega\sqrt{N} \sim |\omega_z|$. In this case squeezing in the light at the cavity output can be found provided that $\Omega\gg \kappa$ when $g\sqrt{N}\sim |\omega_z|$ \cite{Footnote:3}.

\begin{figure}
\subfigure[]{
\includegraphics[width=9cm]{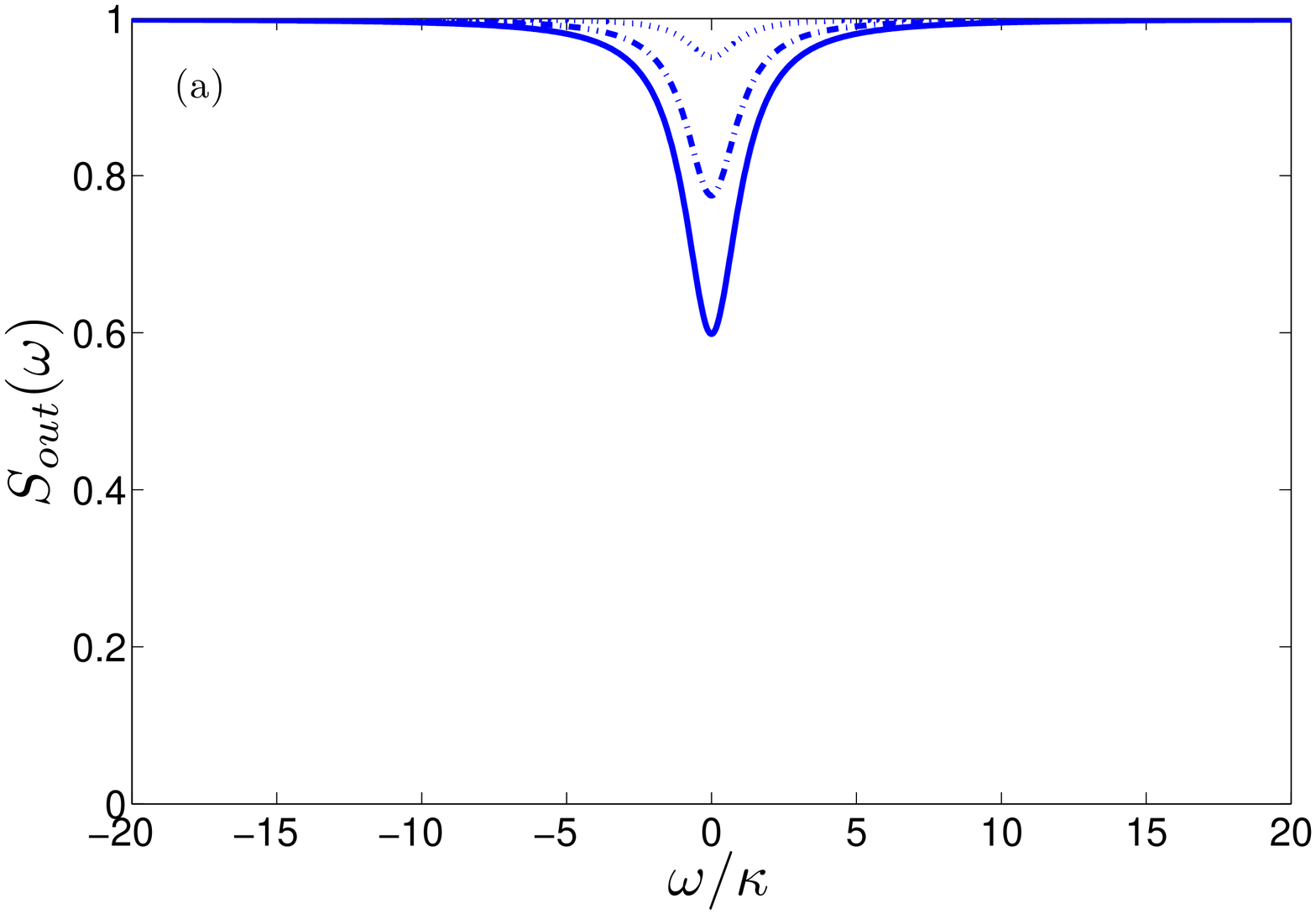}}
\subfigure[]{
\includegraphics[width=9cm]{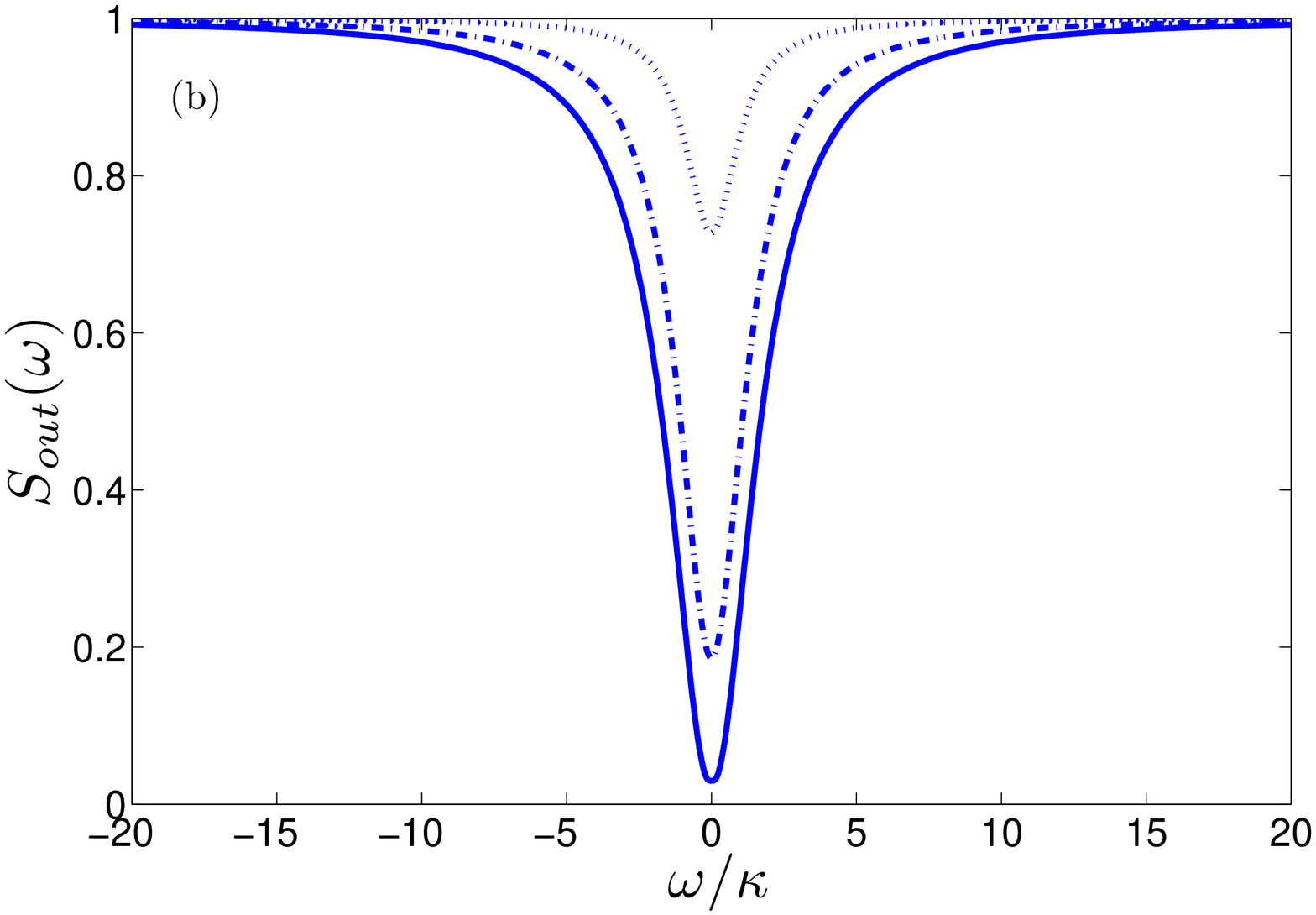}}
\caption[]{(Color online) Squeezing spectrum for the maximum squeezed quadrature when $k=G/2$, $Q'=G/2$ and $Q'\neq Q$ for $\varphi,\phi_L=0$.
The parameters are $\Omega=200\kappa$, $\omega_z=10^3\kappa$ and $N=10,50,100$ atoms (from top to bottom) for (a) $g=4\kappa$ and (b) $g=10\kappa$. The detuning $\delta_c$ is chosen such that $\delta\omega_1$ is zero.
The curves are evaluated from Eq. (\ref{Squeez_spec:0}) by numerically calculating the density matrix of the polariton field for a dissipative dynamics, whose coherent term is governed by the effective Hamiltonian in Eq.~(\ref{Heff_eq:0}). The value $g=4\kappa$ is consistent with the experimental data of Ref.~\cite{Reichel}.
}
\label{Squeez}
\end{figure}

Figures~\ref{Squeez}(a) and (b) display the spectrum of squeezing when the system operates as a parametric amplifier below threshold. Here, one observes that squeezing increases with $N$. Comparison between Fig.~\ref{Squeez}(a) and ~\ref{Squeez} (b) shows that squeezing increases also as the single-atom cooperativity increases (provided the corresponding phase-matching conditions are satisfied and the laser Rabi frequency $\Omega\gg g$). These results agree and extend the findings in Ref.~\cite{Fernandez_07}, which were obtained for an array consisting of 2 atoms.

Let us now focus on the regime when $\varepsilon\ll 1$. Here, the polaritons may be only emitted in pairs into the resonator. In order to characterize the occurrence of these dynamics  we evaluate the
second-order correlation function  at zero time delay in the cavity output defined by \cite{MilburnWalls}
\begin{align}
 g^{(2)}(0)=\frac{\langle a_{\rm out}^{\dag2}a_{\rm out}^2\rangle_{st}}{\langle a_{\rm out}^{\dag}a_{\rm out}\rangle^2_{st}}\,.
\end{align}
Function $g^{(2)}(0)$ quantifies the probability to measure two photon at the cavity output at the same time. Therefore, subpossonian (superpossonian) statistics are here connected to the value of  $g^{(2)}(0)$ smaller (larger) than one, while for a  coherent state $g^{(2)}(0)=1$.

Subpossonian photon statistics at the cavity output can be found as a result of the  dynamics of Eq.~(\ref{Heff_eq:0}). Here, for phase-matching conditions leading to $\nu=0$ and $\alpha\neq 0$, polaritons can only be created in pairs. When the Kerr-nonlinearity is sufficiently large, however, the condition can be reached in which only  two
polaritons can be emitted into the cavity, while emission of a larger number is suppressed because of the blockade due to the Kerr-term. This is reminiscent of the two-photon gateway realized in Ref.~\cite{Kubanek08}, where injection of two photons inside a cavity, pumped by a laser, was realized by exploiting the anharmonic properties of the spectrum of a cavity mode strongly coupled to an atom. In the case analysed in this paper, the anharmonicity arises from collective scattering by the atomic array, when this is transversally driven by a laser. Moreover, we note that the observation of these dynamics requires $\Omega\sqrt{N}\ll |\omega_z|,g\sqrt{N}$ and $|\alpha|>\kappa$, which reduces to the condition $\Omega^2/\omega_z>\kappa$ when $g\sqrt{N}\sim \omega_z$.

\begin{figure}
\centering
\subfigure[]{
\includegraphics[width=8cm]{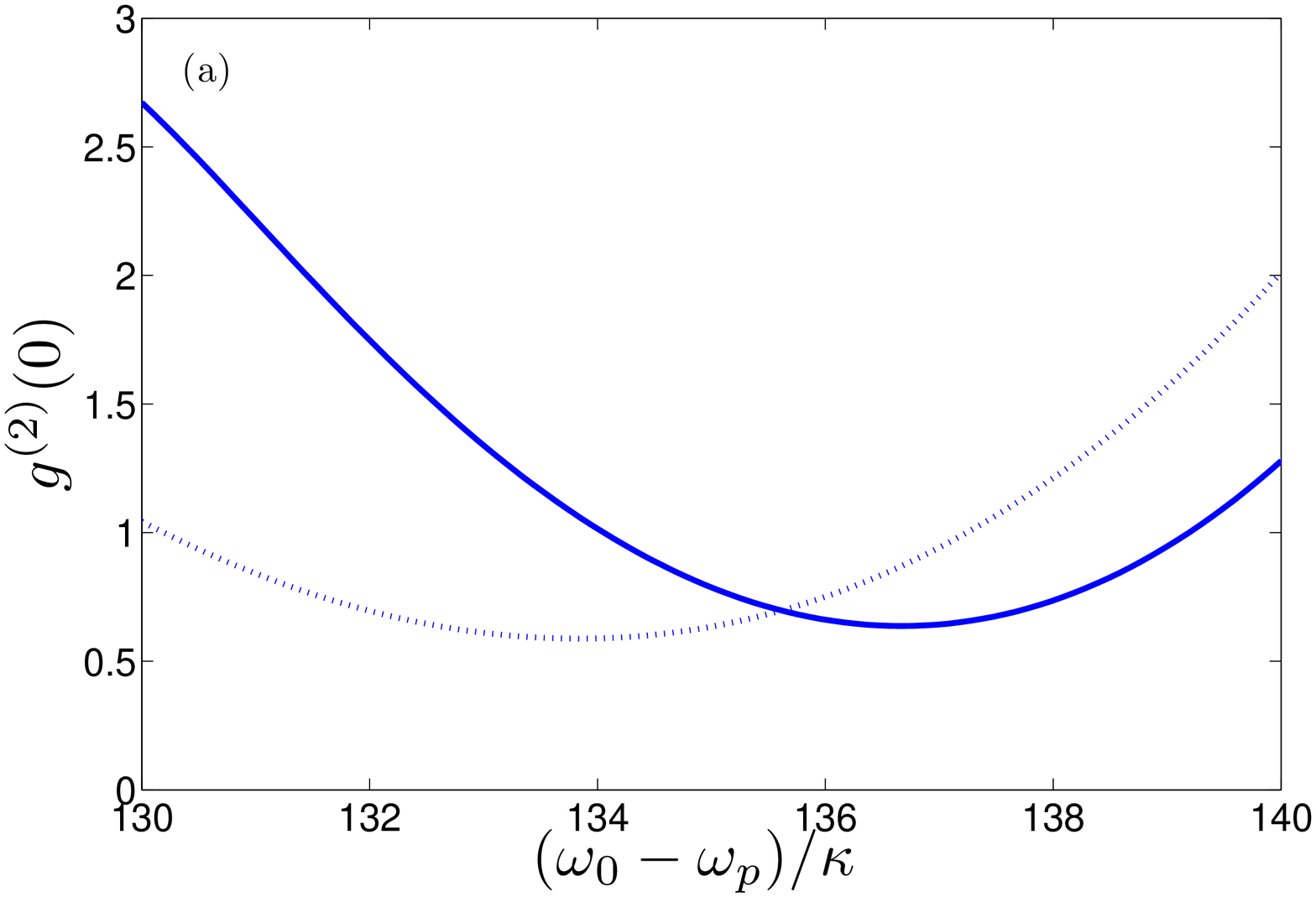}\label{Fig:2phAnti_a}}
\subfigure[]{
\includegraphics[width=8cm]{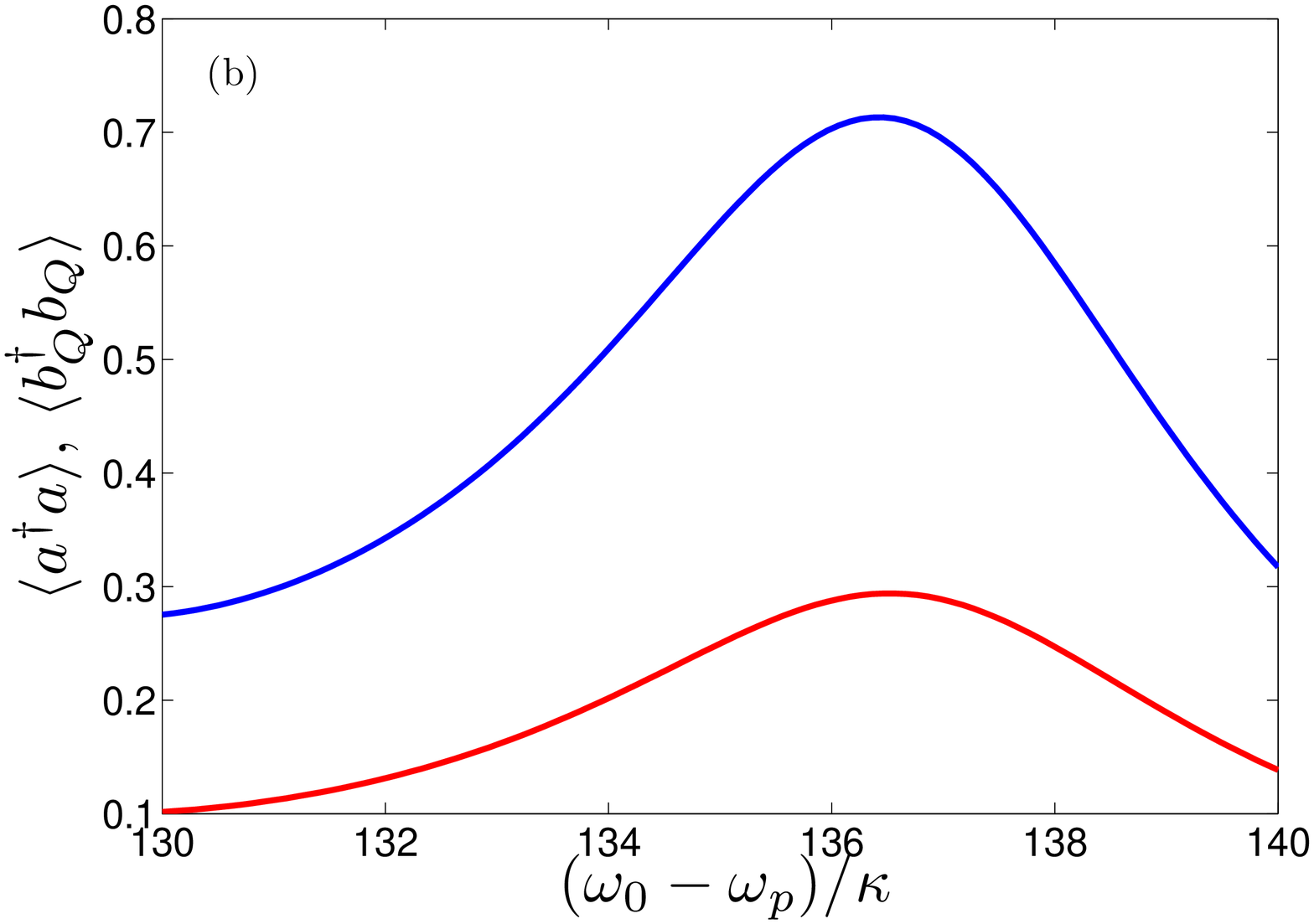}\label{Fig:2phAnti_b}}
\subfigure[]{
\includegraphics[width=8.5cm]{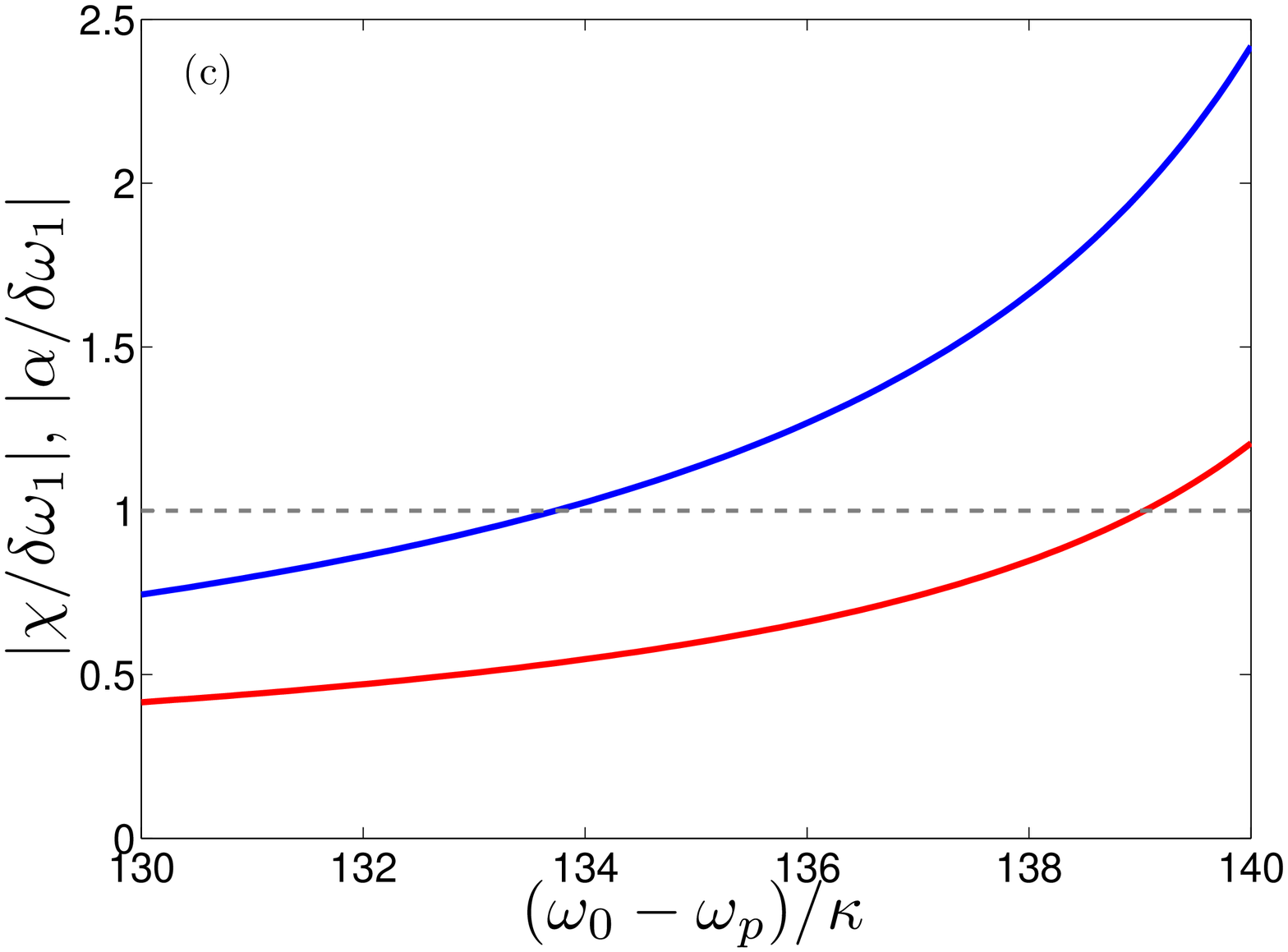}\label{Fig:2phAnti_c}}
\caption[]{(Color online) (a) 
Second order correlation function at zero time delay $g^{(2)}(0)$ versus $\omega_p$ (in units of $\kappa$) when the cavity is solely pumped by inelastic processes  (here, $k=G/2$, $Q'\neq Q,3Q$ and $Q'=G'/2$ for $\varphi,\phi_L=0$). The correlation function is evaluated numerically solving the master equation for the polaritons in presence of cavity decay, with the coherent dynamics given by Hamiltonian~(\ref{Hamilfull}) (solid line) and by Hamiltonian~(\ref{Heff_eq:0}) (dashed line). (b) Corresponding average number of intracavity photons $\langle a^\dag a\rangle$ (blue upper line) and spin wave occupation $\langle b_Q^\dag b_Q\rangle$ (red lower line). (c) Ratios $|\chi/\delta\omega_1|$ (blue upper line) and $|\alpha/\delta\omega_1|$ (red lower line) versus $\omega_p$. The parameters are $g=80\kappa$, $\Omega=30\kappa$, $\omega_z-\delta_c=70\kappa$, and $N=2$ atoms. At the minimum of $g^{(2)}(0)$, $\omega_z\simeq137\kappa$ and $\delta_c\simeq67\kappa$.}\label{Fig:2phAnti}
\end{figure}

Figure~\ref{Fig:2phAnti}(a) displays $g^{(2)}(0)$ as a function of the pump frequency $\omega_p$ for the phase matching conditions giving $\nu=0$ and $\alpha\neq 0$. Function $g^{(2)}(0)$ is evaluated by numerically integrating the master equation with cavity decay, where the coherent dynamics is governed by an effective Hamiltonian which accounts for the effect of both polariton modes and is reported in Eq.~(\ref{Hamilfull}) in the Appendix. Antibunching is here observed over an interval of values of $\omega_p$, about which the cavity mode occupation  has a maximum (blue curve in Fig. ~\ref{Fig:2phAnti}(b)). The maximum corresponds to the value of $\omega_p$ for  which the emission of two polaritons $\gamma_2$ is resonant. Note that the spin-wave excitation, red curve in Fig. ~\ref{Fig:2phAnti}(b), is still sufficiently small to justify the perturbative expansion at the basis of our theoretical model. Figure~\ref{Fig:2phAnti_c} displays the amplitudes $|\chi|$, determining the strength of the Kerr-nonlinearity, and $|\alpha|$, scaling the squeezing dynamics, in units of $|\delta\omega_1|$ and as a function of $\omega_p$. One observes that for the chosen parameters $|\chi|>|\alpha|$. Maximum antibunching is here found when the cavity mean photon number is maximum.

It is important to notice that emission of polaritons in pairs is possible when the collective dipole of the atomic array is driven. For fixed values of $\Omega$ and $g$, we expect that this effect is washed away as $N$ is increased: this behaviour is expected from the scaling of the ratio $\varepsilon$ with $N$. Taking $k=G/2$ and mixing angles $X\ll 1$, for instance, one finds
$\varepsilon\sim \sqrt{N},$
indicating that the strength of the Kerr nonlinearity decreases relative to the coupling $\alpha$ as $N$ grows. This is also consistent with the results reported in Fig. \ref{Squeez}. In this context, the expected dynamics is reminiscent of the transition from antibunching to bunching observed as a function of the number of atoms in atomic ensembles coupled with cavity QED setups~\cite{Hennrich05}.

For the results here presented we have assumed the spontaneous emission rate to be smaller than $\kappa$. In general, the predicted nonlinear effects can be observed in cavities with a large single-atom cooperativity and in the so-called good cavity regime~\cite{Kimble:0}. The required parameter regimes for observing squeezing have been realized in recent experiments~\cite{Reichel}. The parameters required in order to observe a two-photon gateway are rather demanding for the regime in which the atoms are driven well below saturation. Nevertheless, a reliable quantitative prediction with an arbitrary number of atoms would require a numerical treatment going beyond the Holstein-Primakoff expansion here employed.

\section{Conclusion}
\label{Sec:5}

An array of two-level atoms coupling with the mode of a high-finesse resonator and driven transversally by a laser can operate as controllable nonlinear medium. The different orders of the nonlinear responses correspond to different
nonlinear processes exciting collective modes of the array. Depending on the phase-matching condition and on the strength of the driving laser field a nonlinear process can prevail over others, determining the dominant nonlinear response. These dynamics emerge from the backaction of the cavity mode on the scatterers properties, and are enhanced for large single-atom cooperativities. We have focussed on the situation in which the scattering into the resonator is inelastic, and found that at lowest order in the saturation parameter the light at the cavity output can be either squeezed or antibunched. In the latter case, it can either operate as single-photon or two-photon gateway, depending on the phase-matching conditions. Our analysis permits one to identify the parameter regimes, in which a nonlinear-optical behaviour can prevail over others, thereby controlling the medium response. An interesting outlook is whether the considered effects can be used in order to develop turnstile devices, like the one realised in \cite{Dayan}, for quantum networks.

In view of recent experiments coupling ultracold atoms with optical resonators~\cite{Kruse,Slama,Slama_07,Esslinger,Black03,Kinner,Kasevich,Drewsen,Reichel}, these findings show that the coherence properties at the cavity output can be used for monitoring the spatial atomic distribution inside the resonator. Another related question is how the properties of the emitted light depend on whether the atomic distribution is bi- or multiperiodic~\cite{Rist}. In this case, depending on the characteristic reciprocal wave vectors one expects a different nonlinear response at different pump frequency and possibly also wave mixing. When the interparticle distance is uniformly distributed, then coherent scattering will be suppressed. Nevertheless, the atoms will pump inelastically photons into the cavity mode. While in free space the resonance fluorescence is expected to be the incoherent sum of the resonance fluorescence from each atom, inside a resonator one must consider the backaction due to the strong coupling with the common cavity mode.

Finally, in this article we neglect the atomic kinetic energy, assuming that the spatial fluctuations of the atomic center of mass at the potential minima are much smaller than the typical length scales determining the coupling with radiation~\cite{Fernandez_07}. It is important to consider, that when the mechanical effects of the scattered light on the atoms is taken into account, conditions could be found where selforganized atomic patterns are observed~\cite{Domokos02,Kruse,Esslinger,Larson,Vidal10,Domokos10,Goldbart}, which are sustained by nonclassical light. The analysis here presented sets the basis for studies towards this direction.

\subsection*{Acknowledgments}

The authors are grateful to Sonia Fernandez-Vidal, Stefan Rist, Sergey Hritsevitch, and Belen Paredes for stimulating discussions and helpful comments. This work was supported by the European Commission (EMALI MRTN-CT-2006-035369; Integrating project AQUTE; STREP PICC), by the European Science Foundation (EUROQUAM: CMMC), and by the Spanish Ministerio de Ciencia y Innovaci\'on (QOIT, Consolider-Ingenio 2010; QNLP, FIS2007-66944; Ramon-y-Cajal; Juan-de-la-Cierva). G. M. acknowledges the German Research Foundation (DFG) for support (MO1845/1-1). H. H. acknowledges support from the Spanish Ministerio de Ciencia y Innovaci\'on (FPI grant).

\newpage

\appendix

\section{Derivation of the effective Hamiltonian}

>From the general form of the Hamiltonian in Eq.~(\ref{Hgeneral}) for the case in which there is no coherent scattering ($Q'\neq \pm Q$), in the weak excitation limit one can obtain the effective dynamics for the polariton described by $\gamma_1$. We focus on the regime in which $\Omega\sqrt{N}\ll |\omega_z|$.
As we are interested in the dynamics of the mode $b_{Q_s}$ and of the cavity mode $a$, the relevant terms determining their dynamics are given in lowest order by
\begin{eqnarray}\label{Hamilfull}
{\cal H}_{\rm eff}=H_{\rm pump}+\hbar\omega_{Q'}b_{Q'}^{\dagger}b_{Q'}+\hbar \sum_{\sigma=1,2}\omega_\sigma\gamma_\sigma^{\dagger}\gamma_\sigma+{\cal H}'\nn\\
\end{eqnarray}
with
\begin{eqnarray*}
{\cal H}'&&=-\frac{\hbar g}{4\sqrt{N}}\left\{b_{Q'}^{\dagger}b_{Q'}^{\dagger}\left(b_{-Q}{\rm e}^{{\rm i}\varphi}+b_{Q}{\rm e}^{-{\rm i}\varphi}\right)\delta_{Q',G/2}\right.\\
& &+2b_{Q'}^{\dagger}\left(b_{Q}^{\dagger}{\rm e}^{{\rm i}\varphi}+b_{-Q}^{\dagger}{\rm e}^{-{\rm i}\varphi}\right)b_{Q'}\\
& &+\left(b_Q^\dag b_Q^\dag b_Q{\rm e}^{{\rm i}\varphi}+b_{-Q}^\dag b_{-Q}^\dag b_{-Q}{\rm e}^{-{\rm i}\varphi}\right) \\
& &+(1-\delta_{k,G/2})\left[2b_Q^{\dagger}b_{-Q}^{\dagger}b_{-Q}+\delta_{Q,\pm G_0/4}b_{-Q}^{\dagger}b_{-Q}^{\dagger}b_Q \right.\\
& &+2\delta_{3Q,Q'}b_{-Q}^{\dagger}b_{Q'}^{\dagger}b_Q+\delta_{-3Q,Q'}b_{-Q}^{\dagger}b_{-Q}^{\dagger}b_{Q'}\\
& &\left.+\delta_{Q',Q+G/2}b^\dag_{Q'}b^\dag_{Q'}b_{Q} \right]{\rm e}^{{\rm i}\varphi}\\
& &+(1-\delta_{k,G/2})\left[2b_Q^{\dagger}b_{-Q}^{\dagger}b_{Q}+\delta_{Q,\pm G_0/4}b_{Q}^{\dagger}b_{Q}^{\dagger}b_{-Q} \right.\\
& &+2\delta_{-3Q,Q'}b_{Q}^{\dagger}b_{Q'}^{\dagger}b_{-Q}+\delta_{3Q,Q'}b_{Q}^{\dagger}b_{Q}^{\dagger}b_{Q'}\\
& &\left.\left.++\delta_{Q',-Q+G/2}b^\dag_{Q'}b^\dag_{Q'}b_{-Q} \right]{\rm e}^{-{\rm i}\varphi}\right\}a\\
& &-{\rm i}\frac{\hbar\Omega}{2\sqrt{N}}{\rm e}^{-{\rm i}\phi_L}\left\{2b_{Q'}^{\dagger}b_{Q}^{\dagger}b_{Q}+2(1-\delta_{k,G/2})b_{-Q}^{\dagger}b_{Q'}^{\dagger}b_{-Q}\right.\\
& & +\delta_{Q',G/2}b_{Q}^{\dagger}b_{-Q}^{\dagger}b_{Q'}\left[1+(1-\delta_{k,G/2})\right]\\
& &+(1-\delta_{k,G/2})\left[\delta_{3Q,Q'}b_{Q}^{\dagger}b_{Q}^{\dagger}b_{-Q}
+\delta_{-3Q,Q'}b_{-Q}^{\dagger}b_{-Q}^{\dagger}b_{Q}\right.\\
& &\left.\left.+\delta_{Q',Q+G/2}b^\dag_{Q}b^\dag_{Q}b_{Q'}+\delta_{Q',-Q+G/2}b^\dag_{-Q}b^\dag_{-Q}b_{Q'} \right]\right\}\\
& &+ {\rm H.C.}
\end{eqnarray*}
By substituting $b_{Q'}$ with its mean value in Eq.~(\ref{drive}), which corresponds to neglect the backaction on the mode $Q'$ due to the nonlinear coupling, one obtains closed equations of motion for the modes $b_{Q_s}$ and $a$ (where we thereby discard the effect of the nonlinear coupling with the other modes, which are initially empty and which gives rise to higher order corrections). In this limit the effective Hamiltonian~(\ref{Heff_eq:0}) for the polariton $\gamma_1$ is derived provided that the detuning of the laser from the polariton $\gamma_2$ is much larger than the strength of the nonlinear coupling with polariton $\gamma_1$.


\begin{thebibliography}{breitestes Label}
\bibitem{Footnote:0}
This configuration can be achieved by means of an optical lattice trapping the atoms at the minima of the corresponding standing wave, see e.g.~\cite{Kimble}.
\bibitem{Footnote:1}
To be precise, the frequency $\omega_z$ in the denominator of the various coefficients should be replaced by $(\omega_z^2+\gamma^2/4)/\omega_z$, which reduces to $\omega_z$ in the limit $\omega_z\gg\gamma/2$ and which we are going to introduce later on. The coefficients do not include the cavity decay rate in the denominator, under the assumption that it is much smaller than $\omega_z$.
\bibitem{Footnote:3} In this limit, the dependence on the number of atoms is contained in the mixing angle $X$,  Eq. (\ref{angle:X}), and is such that  $\tan X\to 1$ as $N$ is increased.
\end{thebibliography}
\end{document}